\documentclass[dvips]{article}
\usepackage{icrctc07}

\title{Upgrade of the MAGIC Telescope with a Multiplexed Fiber-Optic 2
GSamples/s FADC Data Acquisition system}
\shorttitle{MAGIC MUX FADCs}

\authors{Florian Goebel$^{1}$, Hendrik Bartko$^{1}$, Emiliano Carmona$^{1}$, Nicola Galante$^{1}$,
Tobias Jogler$^{1}$, Razmik Mirzoyan$^{1}$, Jose Antonio
Coarasa$^{2}$, Masahiro Teshima$^{1}$ \\
for the MAGIC collaboration}
\shortauthors{Florian Goebel et al.}
\afiliations{$^1$Max-Planck-Institut f\"ur Physik, Munich\\ 
             $^2$UC San Diego / CERN }
\email{fgoebel@mppmu.mpg.de}

\abstract{In February 2007 the MAGIC Air Cherenkov Telescope for gamma
ray astronomy was fully upgraded with a fast 2~GSamples/s digitization
system. The upgraded readout system uses a novel fiber-optic
multiplexing technique. It consists of 10-bit 2~GSamples/s FADCs to
digitize 16 channels consecutively and optical fibers to delay the
analog signals. A distributed data acquisition system using GBit
Ethernet and FiberChannel technology allows to read out the 100~kByte
events with a continuous rate of up to 1~kHz.}

\begin{document}
\maketitle


\section{Introduction}

MAGIC~\cite{MAGIC} is currently the largest single dish Imaging
Atmospheric Cherenkov telescope (IACT) for high energy gamma ray
astronomy with the lowest energy threshold among existing IACTs. It is
installed at the Roque de los Muchachos on the Canary Island La Palma
at 2200~m altitude and has been in scientific operation since summer
2004. The 17 m diameter parabolic shaped mirror preserves the
time structure of the Cherenkov light signals. The camera is equipped
with 576 photo-multiplier tubes (PMTs). The analog signals are
transfered via optical fibers to the trigger and flash analog to
digital converter (FADC) electronics. Until February 2007 a dual gain
300 MSamples/s 8-bit FADC system~\cite{SiegenFADCs} had been used. This
rather low sampling rate required additional pulses stretching to
ensure proper sampling of the signal.

The gamma ray signals are very short and the PMT response time is very
fast resulting in ~2 ns pulses. A fast readout therefore allows one
to minimize the integration time and thus to reduce the influence of
the background from the light of the night sky (LONS). In addition a precise
measurement of the time structure of the gamma ray signal can help to
reduce the background due to hadronic background events
\cite{MuxPerformance}. The MAGIC collaboration has thus developed a
new, affordable 2 GSamples/s readout system and fully installed it in
February 2007. 

\section{The 2GSamples/s Fiber-Optic Multiplexed readout system}

Fast FADCs are commercially available, but they are prohibitively
expensive and power consuming to read out all the pixels of the MAGIC
camera. To reduce the costs a new 2~GSamples/s read-out system has
been developed and tested~\cite{MuxMirzoyan,MuxTest} at the
Max-Planck-Institut f\"ur Physik in Munich. It uses a novel
fiber-optic multiplexing technique. Multiplexing is possible as the
signal duration (few ns) and the trigger frequency (typically
$\sim$1~kHz) result in a very low duty cycle for the digitizer. The
Fiber-Optic Multiplexing (MUX) technique results in a cost reduction
of about 85\% to using one FADC per read-out channel.

\begin{figure}
  \centering
  \includegraphics[width=0.45\textwidth,angle=0,clip]{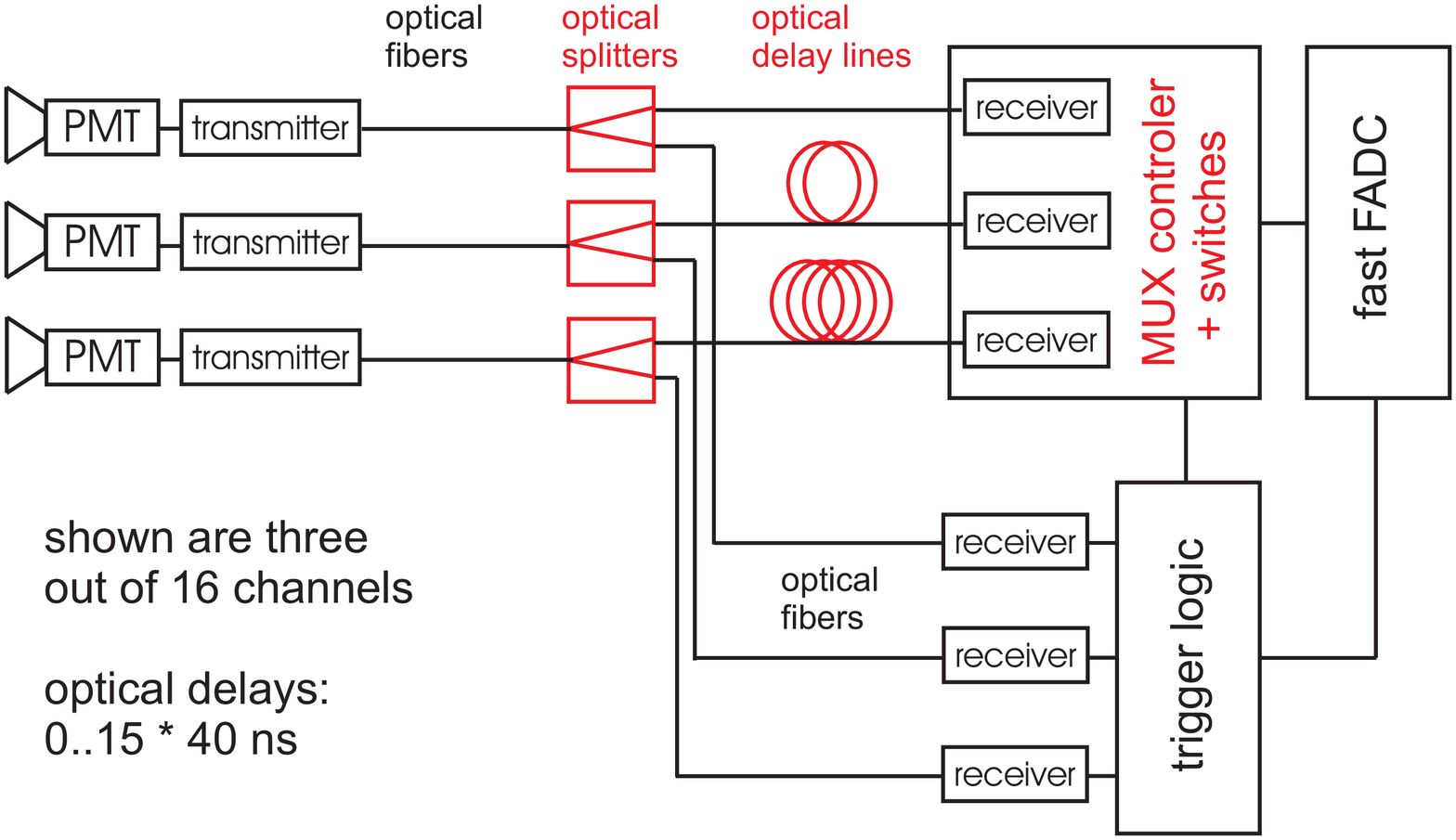}
  \caption{Concept of the fiber-optic multiplexing technique.}\label{fig:MuxConcept}
\end{figure}

The MUX readout system uses a single 2~GSamples/s FADC to digitize 16
read-out channels consecutively. The concept is shown in
Figure~\ref{fig:MuxConcept}. The analog signals are delayed using
optical fibers. Each channel is delayed by 40~ns with respect to the
previous channel. The signals are then electrically multiplexed and
finally sent to the FADC. A trigger signal is generated using a
fraction of the light, which is branched off by fiber-optic light
splitters before the delay fibers.
 
Optical fibers provide very low pulse dispersion and attenuation and
are therefore very well suited for analog signal transmission. Graded
index multimode fibers are used to transmit and delay the 850 nm
wavelength light generated by the VCSEL diodes in the MAGIC
camera. They attenuate the signal by about 2.3 dB for 1 km fiber
length.  GRIN optical splitters with a mode independent 50/50
($\pm3\%$) splitting ratio are used to branch off the trigger signal.
The maximum delay of the fibers after the splitters is 1160 ns to
account for the 40 ns incremental delay for the 16 channels and 560 ns
additional common delay needed to wait for the trigger decision. The
optical delay and splitter modules are assembled in 3U standard 19'' crates.

\begin{figure}
  \centering
  \includegraphics[width=0.5\textwidth,angle=0,clip]{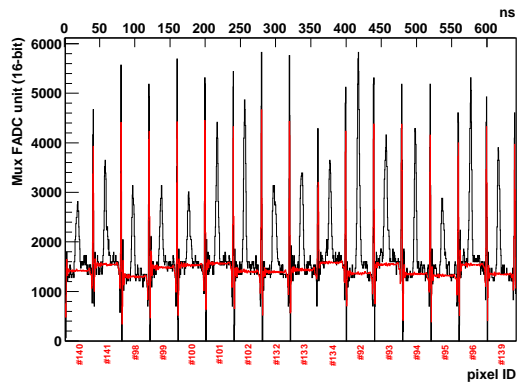}
  \caption{Calibration Event in one FADC Channel with 16 multiplexed
  PMT signals (black) and corresponding pedestal (red).}\label{fig:MuxChannel}
\end{figure}

On the multiplexer boards the optical signals are converted back to
electrical signals using PIN diodes. Fast high bandwidth
MOSFET-switches open for 40 ns one channel at a time. Two switches are
operated in series to reduce the cross talk to about 0.1\%. The
switches are controlled by a digital circuit, which is started by a
trigger signal. Since the control circuit is clocked at 800 MHz the
trigger jitter is 1.25 ns. Finally, the signals are summed in two
active summation stages. In Figure~\ref{fig:MuxChannel} the
multiplexed signal of 16 pixels digitized by one FADC channel is shown. About
10 ns at the borders of every 40 ns window are affected by the
switching process from one channel to the next. These noise peaks have
to be removed by the analysis software.

\section{The 2GSamples/s FADCs}

The multiplexed signals are digitized with DC282 FADC modules produced
by the Acqiris company. The DC282 FADCs feature a 10-bit amplitude
resolution, a bandwidth of 700 MHz, a sampling speed of 2 GSamples/s
and an input voltage range of 1~V. Each DC282 module contains 4 FADC
channels and consumes 60 W power. The FADC intrinsic noise
corresponds to about 1 FADC count. The DC282 FADC modules map the
intrinsic 10-bit FADC data to 16-bit numbers. This mapping compensates
for small FADC intrinsic non-linearities, which are determined during
a special calibration procedure.

The digitized data are stored in an on-board 512 kByte RAM per
channel. The dead time before the next event can be recorded is about
25 $\mu$s. Two FADC boards are inserted in each of the 5 compact PCI
(cPCI) crates. A crate controller PC running Linux reads the data via
the 66 MHz 64 bit cPCI bus. The on-board RAM is logically subdivided
into 3 segments with capacity for 50 events (2.5~kBytes/event). As
soon as 50 events have been written into a segment, new digitized data
is stored in the next segment. The previous segment is then read
out asynchronously via cPCI. A sustainable readout rate of more 1 kHz
with no additional dead time could thus be achieved.

Digital data like trigger bits or the precise time information
provided by a GPS clock is recorded simultaneously with each
event. Four Acromag PMC-DX504 modules installed in two of the cPCI
crates are used to read up to 120 LVDS bits with every trigger. This
data is then read by the crate controller PC and merged with the FADC
data.

\section{The Data Acquisition System}

\begin{figure}
  \centering
  \includegraphics[width=0.5\textwidth,angle=0,clip]{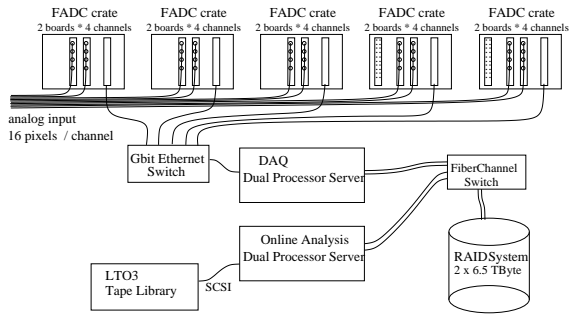}
  \caption{Schematics of MUX FADC data acquisition system.}\label{fig:MuxReadout}
\end{figure}

The schematics of the data acquisition is shown in
Figure~\ref{fig:MuxReadout}. In each of the 5 cPCI crates a data
acquisition program records the digitized data and stores it in a 2000
event deep ring buffer. The data is then sent to the central DAQ
(MUXDAQ), a powerful Dual-Xeon 2.8~GHz server with 8~GB RAM memory
running Linux. The data is transfered over a GBit Ethernet connection
using TCP/IP and a dedicated handshake protocol. A maximum transfer
rate of 100 MByte could be achieved corresponding to a sustainable
trigger rate of 1~kHz.

The main data acquisition program running on MUXDAQ combines the data
of all 5 crates and stores it in a 2000 events deep ring buffer. It
then checks the data integrity, performs a quick signal reconstruction
for online data quality checks and finally saves the data to disk.

The disk storage system was designed to allow high data acquisition
rates and fast simultaneous access to the data for an online data
analysis without disturbing the data acquisition using FiberChannel
and Global File System (GFS) technology.  The storage medium consists
of two EUROstor RAID systems with 16 SATA disks, 500~GBytes each. They
are configured in RAID5. With one spare disk this amounts to
6.7~TBytes storage capacity per RAID system. The RAID systems are
connected via two 2~GBit/s FiberChannel lines and a FiberChannel
switch to MUXDAQ. A maximum rate of 200~MBytes/s could be achieved
writing data to a single RAID system.

In addition to MUXDAQ a second identical server (MUXANA) is connected
to the two RAID systems over the FiberChannel switch. In order to
access the data simultaneously from both servers, GFS has been
installed on the RAID systems. It provides a fast access to the data
without interruption on the data taking, even at these high data
rates.

The MUXANA server is used for quick data analysis, running the
complete MAGIC analysis chain. This allows a quick detection of bright
gamma ray sources. In addition a 36 slot LTO3 tape library is
connected to MUXANA. During one night rawdata of up to 1~TByte are
recorded. This data volume exceeds the bandwidth of the Internet
connection to the MAGIC data center. The rawdata is therefore written
to LTO3 tapes and shipped to the data center. Using compression
1.1~TByte of MAGIC rawdata can be written to one LTO3 tape.
Only the calibrated data which is generated by the quick analysis on
MUXANA is transfered via Internet to the data center and is
available within 24h after data taking.

\begin{figure}
  \centering
  \includegraphics[width=0.45\textwidth,angle=0,clip]{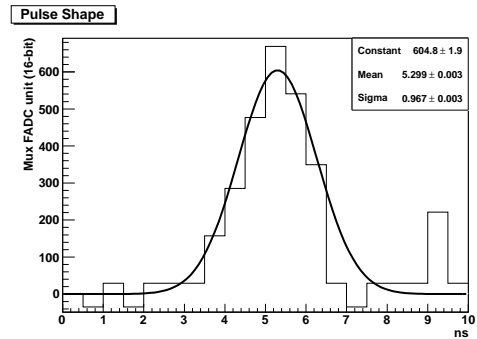}
  \caption{Typical $\sim$5 photoelectron signal (histogram) and mean
  scaled pulse shape of a single photoelectron signal.}\label{fig:MuxPulse}
\end{figure}

\section{System Performance}

Since the installation in February 2007 the MUX FADC system has been
taking data smoothly. A sustainable data acquisition rate of up to
100~MBytes/s corresponding to a trigger rate of 1~kHz has been
achieved.

In Figure~\ref{fig:MuxPulse} a signal corresponding to $\sim$5 photoelectrons
is shown. A PMT afterpulse signal has been selected to study the
response of the whole signal chain to a maximally short input
signal. A pulse width of FWHM of 2.3~ns is obtained. 

\begin{figure}
  \centering
  \includegraphics[width=0.45\textwidth,angle=0,clip]{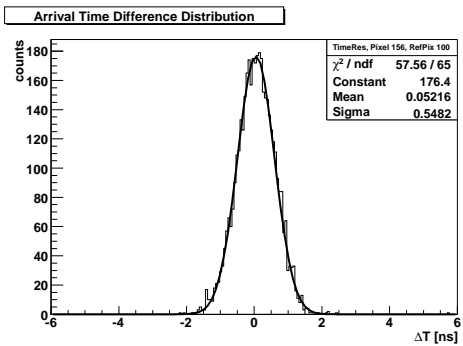}
  \caption{Distribution of the arrival time difference between two
  camera pixels for calibration events.}\label{fig:MuxTimeRes} 
\end{figure}

In order to study the time resolution of the whole signal chain,
calibration events have been used, which illuminate the camera
uniformly with short (few ns) light pulses. In
Figure~\ref{fig:MuxTimeRes} the distribution of the arrival time
difference between two camera pixels is shown for calibration
events. Two random inner camera pixels have been chosen for which the
mean light pulse intensity corresponds to $\sim$35 photoelectrons. The
arrival time is calculated with the simple algorithm $t = \sum_i t_i
q_i / \sum_i q_i$, where $i$ is the FADC slice index, and $t_i$ and
$q_i$ are the time and the charge of slice $i$. The RMS of the
distribution is 550 ps corresponding to a time resolution of 390 ps.

\begin{figure}
  \centering
  \includegraphics[width=0.45\textwidth]{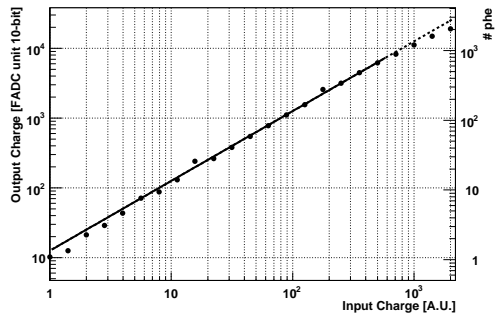}
  \caption{Linearity of the FADC signal as a function of the input pulse
    area.}\label{fig:MuxLinearity}
\end{figure}

The signal gain is currently adjusted such that a single photoelectron
produces a mean signal with an amplitude of $\sim$2 FADC counts and
an area of $\sim$9 FADC counts. It thus roughly corresponds to a 3
sigma FADC noise level. Saturation effects of the multiplexer
board start to become noticeable at signals corresponding to about 800
photoelectrons, thus leading to a linear dynamic range of 800.
In Figure~\ref{fig:MuxLinearity} the linearity of the reconstructed
FADC pulse is shown as a function of the input pulse area.
 
The potential and performance of the fast 2~GSamples/s MUX FADC system
to reduce background and to increase the sensitivity of the MAGIC
telescope are addressed in\cite{MuxPerformance}.

\section{Acknowledgments}

We would like to thank the German MPG for special support for the MUX
FADC upgarde project and the IAC for excellent working conditions. The
support of the German BMBF and MPG, the Italian INFN and the Spanish
CICYT, the Swiss ETH and the Polish MNiI is gratefully acknowledged.

\bibliography{library}
\bibliographystyle{plain}

\end{document}